\documentclass[letterpaper, 10 pt, conference]{ieeeconf}  %

\IEEEoverridecommandlockouts                              %

\usepackage{amsmath,amsfonts,amssymb}
\usepackage{bbm}
\usepackage{algorithmic}
\usepackage{algorithm}
\usepackage{array}
\usepackage[caption=false,font=footnotesize,labelformat=simple]{subfig}

\usepackage{textcomp}
\usepackage{stfloats}
\usepackage{url}
\usepackage{verbatim}
\usepackage{graphicx}
\usepackage{cite}
\usepackage{booktabs}
\usepackage[dvipsnames]{xcolor}

\usepackage{url}
\usepackage{xcolor}
\usepackage{sj_mysymbol}

\title{\LARGE \bf
A Performative Control Approach to Day-to-Day Traffic Signal Retiming 
\thanks{*Corresponding author: Zhongyuan Zhao.}%
\thanks{This work was supported by NSF under award CCF-2340481.}%
\thanks{Emails: \{\texttt{shuchang.jiang}, \texttt{xinwu.qian}, \texttt{segarra}, \texttt{zhongyuan.zhao}\}@rice.edu }
}

\author{Shuchang Jiang, Xinwu Qian, Santiago Segarra, and Zhongyuan Zhao$^*$\\
\textit{Rice University}, Houston, TX, USA
}

\begin{document}

\maketitle
\thispagestyle{empty}
\pagestyle{empty}

\begin{abstract}
Conventional fixed-time traffic signal optimization and recent day-to-day signal retiming generally model route adaptation under prescribed route-choice behavior and origin--destination (OD) demand, while neglecting the fact that travelers' behavior can be performative to the signal control policy. 
A signal timing policy learned from historical traffic observations must therefore account for how its own deployment changes the distribution on which it will be evaluated in the future.
We formulate this problem as performative signal control under observed traveler responses and characterize performative stability of the coupled policy--demand system.
We propose incremental day-to-day retiming using the local policy gradient at the current operating point, estimated offline through simulation-based zeroth-order optimization without an explicit day-to-day demand-response model.
This gradual approach provides a feasible path for policy and demand to co-evolve toward stability.
Experiments on a downtown Denver network show that our approach improves congestion, throughput, and stability, generalizes across different demand levels, and remains effective when the offline simulator contains mismatched travel demand and routing behaviors.
\end{abstract}


\section{Introduction}

Traffic signal optimization can reduce urban congestion without costly road expansion, bringing substantial mobility, economic, and environmental benefits. Existing actuated and real-time traffic signal control methods~\cite{levin2023maxpressure,liang2025survey} rely on continuous, fine-grained traffic-state sensing and communication to adjust signals in real time.
Such infrastructure remains costly to deploy and maintain at urban scale~\cite{mladenovic2012}.
In contrast, aggregate traffic observations over longer intervals are more readily available from existing detectors and increasingly from mobile-phone data, for which even 2--3\% penetration can provide accurate traffic state measurements~\cite{herrera2010}.
Yet most urban signals still use fixed-time control, with retiming performed only every 3--5 years~\cite{wang2024}, too infrequently to keep pace with changes in traffic demand. This gap motivates day-to-day signal retiming~\cite{hu1997day,cantarella_watling_2015_day2day,xiao2015combined,smith2015traffic,huang2016}: timing plans remain fixed within an operating period, such as the morning peak, but are updated across days using aggregate traffic observations. These updates allow timing plans to adapt to changing demand while retaining fixed-time operation on existing signal infrastructure.

Signal-timing optimization has been studied extensively under known travel demand and traveler behavior.
At the intersection level, Webster's method~\cite{webster1958traffic} treats the observed approach flows as given and determines signal timings accordingly.
At the network level, combined traffic assignment and control (CTAC)~\cite{yang1995traffic} jointly optimizes signals and route choices, typically through a Wardrop user equilibrium~\cite{wardrop1952} for a prescribed origin-destination (OD) demand.
Simulation-based optimization captures traffic responses implicitly through repeated simulation and directly searches the policy space using derivative-free methods~\cite{osorio2013} or high-dimensional Bayesian optimization such as TuRBO~\cite{gong2025turbo}.
In general, these methods optimize a timing policy for a prescribed demand and its associated traffic state.
In reality, however, travelers often adapt route choices and travel demand based on deployed policy and recent experiences~\cite{cantarella_watling_2015_day2day}, so the policy itself changes the operating condition under which it was optimized. This policy-dependent distribution shift, known as the performative effect~\cite{perdomo2020performative,hardt2025performative,zhang2026performative}, can invalidate a policy optimized for the original traffic condition and motivates explicitly accounting for day-to-day adaptation.

Existing literature on day-to-day retiming models signal control and route choice as a coupled dynamical process.
In~\cite{xiao2015combined}, signal timing is updated by Webster's method based on previous-day observations, and the stability of the resulting route-choice control dynamics is analyzed.
In~\cite{smith2015traffic}, green time is incrementally reallocated according to observed traffic conditions, and convergence of the coupled user--control adaptation is established under suitable behavioral assumptions.
In~\cite{huang2016}, repeated observations are used to correct model bias and response sensitivity, enabling incremental convergence toward the realized anticipatory optimum.
These studies demonstrate the value of observation-driven incremental control, but capture only route adaptation behaviors under fixed OD demands.

In this work, we take one step further by considering the performative effects of signal control on both how travelers choose routes and whether they enter a signalized network, allowing deployed policies to reshape traffic assignment and network demand.
Under this behavioral regime, as shown in our experiments, even a fixed timing policy can exhibit substantial day-to-day fluctuations in demand and traffic cost, while repeatedly solving for an end-state optimum does not by itself provide a stable path for the coupled evolution of travelers and control.
We instead employ incremental retiming based on the local policy gradient at the current operating point, providing a gradual path for policy and demand to co-evolve toward performative stability.

The contributions of this work are threefold.
First, we formulate fixed-time signal retiming as a performative control problem and characterize performative stability, where travelers change their travel decisions with respect to experienced costs and the timing policy is stationary for that demand.
Second, we develop a simulation-based zeroth-order gradient framework that estimates policy updates offline under observed OD demand, enabling gradual retiming without an explicit model of the day-to-day demand response.
Third, numerical experiments show that the proposed approach stabilizes the joint evolution of timing and demand, improves network performance over static and responsive-control baselines, remains robust to OD-observation noise and simulator mismatch, and generalizes across substantially different demand levels.

\section{System Model and Problem Formulation}\label{sec:system}
We study a static road network $\ccalG$ with $p$ controllable signalized intersections, on which the fixed timing policy for an operating interval (e.g., morning peak) on day $t$ is denoted as $\bbx(t)\in\mathcal Z\subseteq\mathbb R^p$ (e.g., green split), where $\mathcal Z$ represents the feasible policy set.
The traffic system response of day $t$ can be represented as the following function:
\begin{equation}\label{E:system}
    \ccalV(t), C(t)=\Phi_{\ccalG}(\bbx(t),F(t)),
\end{equation}
where vector $F(t)\in\mathbb R_{+}^{K}$ collects $K$ pairs of the OD demand on day $t$, $\ccalV(t)$ is the set of served trips, and $C(t)\in\reals_+^{K}$ is the realized OD cost vector (e.g., average OD timeloss). 
Function $\Phi_{\ccalG}(\cdot)$ captures within-day route adaptation of the traveling population and the resulting traffic execution.
Hereafter, we omit the static road network $\ccalG$ for simplicity.

For notational convenience, we also denote the OD-cost vector as a function $C(t)=C(\bbx(t),F(t))$,
of which the $k$th element $C_k(t)$ or $C_k(\bbx(t),F(t))$ stands for the average realized travel cost of OD pair $k\in\{1,\dots,K\}$.
Similarly, let $\ccalV_k(t)=\ccalV_k(\bbx(t),F(t))$ denote the set of served trips for OD pair $k$, such that $\ccalV(t) = \bigcup_{k=1}^{K}\ccalV_k(t)$.

\subsection{Traveler Adaptation}\label{sec:system:behavior}

Travelers adapt their travel decisions in response to the signal timing policies.
Within the same day, self-interested travelers may adjust their routes according to traffic conditions, which is captured implicitly by $\Phi_{\ccalG}$ in~\eqref{E:system}.

At the beginning of each day $t$, the traveling population perceives the OD cost based on past experience:
\begin{equation}\label{E:perceived_cost}
    c_k(t+1)=f_c\bigl(c_k(t),C_k(t)\bigr),
\end{equation}
where $ c_k(t)$ is the perceived cost for OD pair $k$ on day $t$, and vector $\bbc(t)=[c_k(t)]_{k\in\{1,\ldots,K\}}^\top$.
We assume that $f_c$ is continuous, nondecreasing in its realized-cost argument, 
and consistent in the sense that $f_c(c,c)=c$.

Travelers then decide whether to take the automobile trips, making travel demand a function of the perceived cost:
\begin{equation}\label{E:demand_response}
    F_k(t)=f_d\bigl(c_k(t)\bigr).
\end{equation}
We assume that $f_d$ is continuous, convex, nonincreasing in perceived travel cost, 
so that higher perceived cost does not increase demand and its sensitivity remains bounded.

The exact forms and parameters of functions $\Phi_{\ccalG}$, $f_c$, and $f_d$ are unknown to the controller.
However, we assume that $ \ccalV(t), C(t), F(t) $ in~\eqref{E:system} can be estimated with finite accuracy from traffic sensing and estimation algorithms~\cite{cascetta1993}, and that $\Phi_{\ccalG}$, including its route-adaptation effects, can be approximated by a traffic simulator.

\subsection{Day-to-Day Behavioral Data Generation}\label{sec:system:data}

For numerical evaluation, we generate the environment by instantiating the otherwise unknown day-to-day behavioral mappings in Sec.~\ref{sec:system:behavior} with specific models.
The perceived OD cost in~\eqref{E:perceived_cost} follows an exponential moving average (EMA)~\cite{hunter1986}:
\begin{equation}\label{E:cost}
    c_k(t+1)
    =
    (1-\gamma)c_k(t)
    +
    \gamma C_k(t),
    \qquad 0<\gamma\leq1,
\end{equation}
where smaller $\gamma$ weighs more in past travel experiences over the most recent realized cost.
The demand on an OD pair in~\eqref{E:demand_response} follows the exponential elastic demand model~\cite{ryu2014}:
\begin{equation}\label{E:exp_demand}
    F_k(t)
    =
    F_k^{\min}
    +
    \bigl(F_k^{\mathrm{ref}}-F_k^{\min}\bigr)
    \exp\!\left\{
    \beta_k\bigl(c_k^{\mathrm{ref}}-c_k(t)\bigr)
    \right\},
\end{equation}
where $F_k^{\min}$ is the inelastic demand, $F_k^{\mathrm{ref}}$ is the reference demand associated with $c_k^{\mathrm{ref}}$, and $\beta_k>0$ controls demand sensitivity to perceived cost.
These models are used only to generate the day-to-day traffic environment; the controller observes their resulting demand and traffic outcomes without access to their forms or parameters.

\subsection{Problem Formulation}
Following the individual-benefit interpretation of consumer surplus~\cite{boyles2025transportation}, we define the network-level objective function \(J(t)\) as total social surplus (TSS), measured relative to the outside option:
\begin{equation}\label{E:abstract_tss}
    J(t)
     \!=\! J(\bbx(t),F(t))
     =\! \sum_{k=1}^{K} \;
    \mathbb{E}\!\left[\frac{|\ccalV_k(t)|}{T_h}\left(
         b_k(t)
        \!-\!
         C_k(t)\right)
    \right],    
\end{equation}
where \(b_k(t)>0\) and \(C_k(t)\) denote the average economic benefit and travel cost of an automobile trip for OD pair \(k\), both measured relative to an outside option such as transit or foregoing the trip. The ratio \(|\mathcal V_k(t)|/T_h\) gives the realized total network throughput. Improving TSS hence requires balancing the benefits of increasing network throughput against managing the induced congestion from higher traffic.  

The control problem then seeks to maximize TSS by updating daily signal timing policy, using up-to-date traffic observations. On day \(t\), the policy \(\bbx(t)\) and demand \(F(t)\) jointly determine network traffic outcomes, which also shapes the  local experienced costs that changes subsequent demand through the latent behavioral mappings in \eqref{E:perceived_cost}–\eqref{E:demand_response}. 
Consequently, a timing policy optimized for currently observed demand can change the demand under which it is subsequently evaluated. 
An ideal controller is supposed to lead this policy-dependent evolution towards a performatively optimal policy $\bbx^*$ that maximizes the TSS under the equilibrium demand induced by  itself:
\begin{equation}\label{E:performative-opt}
\bbx^*
\in
\argmax_{\bbx\in\mathcal Z}
J\bigl(\bbx,F^\star(\bbx)\bigr)\;.
\end{equation}
In~\eqref{E:performative-opt}, $F^\star(\bbx)$ denotes the equilibrium demand induced by static policy $\bbx$, which satisfies, for all $
k\in\{1,\dots,K\}$
\begin{subequations}\label{E:stationary-demand}
\begin{align}
F_k^\star(\bbx)
& =
f_d\left(
 c^\star_k(\bbx)
\right), \\
c^\star_k(\bbx)&=f_c\left(c^\star_k(\bbx),C_k\bigl(\bbx,F^\star(\bbx)\bigr)\right).
\end{align}
\end{subequations}

In real-world applications, directly optimizing the above coupled process is challenging because \(J(\boldsymbol{x},F)\) is induced by the black-box traffic response in~\eqref{E:system}, while the day-to-day behavioral mappings are unknown to the controller. 
To address this issue, one possible approach is repeated risk minimization, which repeatedly optimizes the policy using the observations generated by previous policy deployments. 
Under suitable conditions, this procedure is known to converge to a performatively stable solution~\cite{perdomo2020performative}: a policy that is optimal for the demand it induces. 
Such a solution need not be performatively optimal, since performative optimality additionally accounts for how demand changes when comparing alternative policies. 
In this study, we pursue an incremental adaptation approach, seeking a self-consistent operating point at which signal timing is first-order stationary for the demand it induces. Specifically, our approach uses an approximate traffic simulator to evaluate timing changes conditional on the currently estimated OD demand and guide incremental policy updates. Subsequent behavioral responses are then incorporated through new traffic observations, allowing signal timing and demand to evolve jointly without requiring an explicit model of the day-to-day demand response. 
The next section presents the update procedure and characterizes its stationary target.

\section{Simulation-Based Zeroth-Order Performatively Stable Control}\label{sec:solution}

We seek to approximate~\eqref{E:performative-opt} through local gradient-based optimization.
Although $J(\bbx,F)$ needs not be convex, it can be assumed to be smooth with respect to the signal-timing policy $\bbx$ and traffic demand $F$, since sufficiently small perturbations of either generally lead to gradual rather than discontinuous changes in aggregate network performance. 
The policy gradient for performative optimal control is:
\begin{equation}
\begin{aligned}
\nabla_{\bbx}J\bigl(\bbx,F^\star(\bbx)\bigr)
&=
\underbrace{
\left.
\nabla_{\bbx}J(\bbx,F)
\right|_{F=F^\star(\bbx)}
}_{\substack{\text{fixed-demand timing effect}\\
\text{including adaptive routing}}}
\\
&\quad+
\underbrace{
\left(
\frac{\partial F^\star(\bbx)}
{\partial \bbx}
\right)^{\!\top}
\left.
\nabla_F J(\bbx,F)
\right|_{F=F^\star(\bbx)}
}_{\text{long-run induced-demand effect}}.
\end{aligned}
\label{E:performative-gradient}
\end{equation}

The first term captures the timing effect under fixed OD demand, including within-day route adaptation, while the second captures the effect through policy-induced changes in stationary demand.
Because the behavioral mappings governing $F^\star(\bbx)$ are unknown, its sensitivity $\partial F^\star(\bbx)/\partial\bbx$ is generally unavailable.
We therefore omit the second term in~\eqref{E:performative-gradient} and seek only \textit{performative stability} by incrementally optimizing the policy for the currently observed demand~\cite{cascetta1993} while allowing demand to adapt across days.

Given the estimated demand $\widehat F(t)$, we use an offline traffic simulator and zeroth-order optimization to estimate the current-demand policy gradient $\nabla_{\bbx}J(\bbx,F)$.
A single traffic realization provides a noisy observation $\widehat J(\bbx(t),F(t))$ of the true TSS objective $J(\bbx(t),F(t))$, where
\begin{equation}\label{E:observed_TSS}
    \widehat J(\bbx(t),F(t))
    = \sum_{k=1}^{K} \;
    \frac{|\ccalV_k(t)|}{T_h}\bigl(
         b_k(t)
        -
         C_k(t)\bigr),   
\end{equation}
and we assume that the expected simulator response provides a sufficiently accurate local approximation of the real-system objective for policy optimization.
The policy $\bbx$ is updated through projected gradient ascent:
\begin{equation}\label{E:proj_ascent}
    \bbx(t+1)
    =
    \Pi_{\mathcal Z}
    \left[\bbx(t)+\eta_t \widehat{\nabla_{\bbx(t)}} J\bigl(\bbx(t), F(t)\bigr)\right],
\end{equation}
where $\eta_t>0$ is the update step size and $\Pi_{\mathcal Z}$ denotes the Euclidean projection onto the feasible signal-timing set $\mathcal Z$, subject to regulatory constraints of minimum and maximum green time.
The projection  $\bbx=\Pi_{\mathcal Z}(\bbz)$ is expressed as
\begin{subequations}\label{E:projection}
\begin{align}
    \bbx &= \argmin_{\tilde\bbx} \frac{1}{2}|| \tilde \bbx - \bbz ||^2_2 \\
    \text{s.t.}\quad & x_j^{\min}\leq\tilde x_j \leq x_j^{\max},\quad \forall j\in\{1,\dots,p\}\;.
\end{align}
\end{subequations}
Problem~\eqref{E:projection} is convex and can be solved efficiently.

The procedure of day-to-day performatively stable control is illustrated in Algorithm~\ref{A:day-to-day}, where we rely on observed traffic and estimated demand rather than explicit demand-response models. 
\begin{algorithm}[t]
\caption{Day-to-Day Performatively Stable Control}
\label{A:day-to-day}
\begin{algorithmic}[1]
\REQUIRE Initial policy $\bbx(1)\in\mathcal Z$, ZOO sample size $M$.
\STATE Initialize perceived cost $c(1)$.
\FOR{$t=1,2,\ldots$}
    \STATE Observe traffic realization from $\Phi_{\ccalG}(\bbx(t),F(t))$ and estimate OD demand $\widehat F(t)$.
    \STATE Estimate $\widehat{\nabla_{\bbx}}J(\bbx(t),\widehat F(t))$
    with $M$-sample ZOO~\eqref{E:general_zoo}.
    \STATE Update $\bbx(t+1)$ based on~\eqref{E:proj_ascent} and~\eqref{E:projection}.
    \STATE Update $\bbc(t+1)$ according to~\eqref{E:cost}.
    \STATE The population responds to $\bbc(t+1)$ according to the demand-response mapping $f_d$, producing $F(t+1)$.
\ENDFOR
\end{algorithmic}
\end{algorithm}

\subsection{Zeroth-Order Policy Gradient Estimation}

Despite recent progress in first-order optimizations based on surrogate models of non-differentiable networked systems~\cite{zhao2022graph,li2024glance,zhao2025asilomar}, we rely on zeroth-order gradient estimation due to the lack of capable surrogate models to learn and predict the empirical behaviors of traffic systems. 

For each day $t$, we generate $M$ random policy perturbations
$\bbn_m\sim\mathcal N(\bbzero,\sigma^2\bbI)$ and evaluate the perturbed
policies in offline simulation under the same estimated OD demand
$\widehat F(t)$. 
For current $\bbx(t),F(t)$, omitting $t$ for notational simplicity, a basic multi-point zeroth-order estimate of the conditional policy gradient is \cite{liu2020primer}
\begin{equation}\label{E:zoo_gradient}
\widehat{\nabla_{\bbx}J}
=
\frac{1}{M\sigma^2}
\sum_{m=1}^{M}
\left[
\widehat J\bigl(\bbx+\bbn_m,\widehat F\bigr)
-
\widehat J\bigl(\bbx,\widehat F\bigr)
\right]\bbn_m .
\end{equation}

More generally, let the chosen offline simulator collects $M$ query data around the current policy $\bbx$
\begin{equation}
\ccalD =
\left\{
\left(
\bbx \!+\!\bbn_m,
\widehat J(\bbx \!+\! \bbn_m,\widehat F),
\widehat{\bbw}(\bbx \!+\! \bbn_m,\widehat F)
\right)
\right\}_{m=1}^{M}
\end{equation}
{where vector $\widehat \bbw\in\reals_+^{p}$ collects the average waiting time at controllable intersections.}
Let an advanced zeroth-order gradient estimator be represented abstractly as
\begin{equation}\label{E:general_zoo}
    \widehat{\bbg}
    =
    \ccalA_{\mathrm{ZOO}}
    \left(\ccalD,\ccalH\right),
\end{equation}
where $\ccalH$ denotes optional historical query information.  
The multi-point estimator in~\eqref{E:zoo_gradient} is a basic stateless example, while the framework is agnostic to the choice of $\ccalA_{\mathrm{ZOO}}$.

In our implementation, we choose ZoRD~\cite{shu2023zerothorder} as $\ccalA_{\mathrm{ZOO}}$, which fits a Gaussian-process (GP) surrogate to available query history and obtains gradient estimates together with their posterior uncertainties.
To improve spatial credit assignment, we augment the network-wide TSS objective with junction-level waiting-time regularization:
\begin{equation}\label{E:obj_zord}
    \Tilde{J} = \widehat J - w||\widehat \bbw||_1\;,\qquad \text{where}\; w\geq 0.
\end{equation}
To exploit the uncertainty weighted mechanism in ZoRD, we fit a global GP to $\widehat J$, while using independent local GPs to model the elements of $\widehat\bbw$.
Their gradient estimates are fused coordinate-wise according to posterior uncertainty:
\begin{equation}
{\bbg}_j
=
\frac{
\bbu_j^{\mathrm{r}}\cdot
\nabla_{\bbx_j}\mu^{\mathrm{o}}\big(\widehat{J}\;\big)
- 
w\cdot\bbu_j^{\mathrm{o}}\cdot 
\nabla_{\bbx_j}\mu^{\mathrm{r}}(||\widehat \bbw||_1)
}{
\bbu_j^{\mathrm{o}}
+
\bbu_j^{\mathrm{r}}
}\;,
\label{E:fused_gradient}
\end{equation}
where $\mu^{\mathrm{o}},\mu^{\mathrm{r}}$ denote the corresponding GP posterior means and
$\bbu^{\mathrm{o}},\bbu^{\mathrm{r}}$ the posterior variances of their gradient estimates.
Thus, more reliable gradient components receive larger weights.
Within $\ccalA_{\mathrm{ZOO}}$, this raw fused gradient $\bbg$ is further processed using its uncertainty and optimization history through uncertainty-adaptive multiple gradient steps and Adam, yielding the final update gradient $\widehat{\bbg}$ used in~\eqref{E:proj_ascent}.

In Section~\ref{sec:results}, the effectiveness of our method is evaluated under real-sim gap, i.e., the traffic demand and route adaptation behaviors in offline simulations are inaccurate compared to $\Phi_{\ccalG}$ the actual traffic system under control.

\subsection{Performative Stability}

We characterize the performatively stable operating point in day-to-day control as follows. 
Let $\bbx_{\rm PS}$ denote a stable static signal-timing policy and let
$F_{\rm PS}=F^*(\bbx_{\rm PS})$ denote its induced equilibrium demand. In the idealized mean system, a performatively stable stationary point satisfies
\begin{subequations}\label{E:pstable}
\begin{align}
F_{\rm PS}
&=
f_d\left(
\mathbb E[C(\bbx_{\rm PS},F_{\rm PS})]
\right),\\
0
&\in
-\nabla_{\bbx}J(\bbx_{\rm PS},F_{\rm PS})
+
N_{\mathcal Z}(\bbx_{\rm PS}),
\end{align}
\end{subequations}
where $N_{\mathcal Z}(\bbx_{\rm PS})$ denotes the normal cone to the feasible policy set at $\bbx_{\rm PS}$. The first condition requires the travel demand to be consistent with the expected traffic cost induced by the policy, while the second requires the policy to be first-order stationary for the demand that it induces.

In practice, however, the day-to-day signal-timing policy $\bbx(t)$ is continuously updated by noisy policy gradient estimated from noisy observations and offline simulations of a stochastic underlying traffic system. 
Therefore, we consider the controller has achieved performative stability if the day-to-day fluctuations of policy and demand are bounded around some equilibrium points $(\bbx_{\rm PS},F_{\rm PS})$:
\begin{subequations}\label{E:pstable-neighborhood}
\begin{align}
\limsup_{t\rightarrow\infty}
\mathbb E\!\left[
\|\bbx(t)-\bbx_{\rm PS}\|_2^2
\right]
&\leq \delta_x,\\
\limsup_{t\rightarrow\infty}
\mathbb E\!\left[
\|F(t)-F_{\rm PS}\|_2^2
\right]
&\leq \delta_F,
\end{align}
\end{subequations}
where $\delta_x,\delta_F\geq0$ characterize the steady-state fluctuations caused by stochasticity and estimation errors.

\begin{table}[!t]
\centering
\caption{Experimental configuration.}
\label{tab:experiment_settings}
\footnotesize
\renewcommand{\arraystretch}{0.95}
\setlength{\tabcolsep}{3pt}
\begin{tabular}{@{}p{0.25\columnwidth}p{0.68\columnwidth}@{}}
\textbf{Name} & \textbf{Settings} \\
\toprule

Online plant  & SUMO, $\rho=1.0$, $F(t)$ evolves by \eqref{E:system}, \eqref{E:cost}, \eqref{E:exp_demand}; \\

(under control) & Simulation time: warm-up 900s, main 3600s; \\
& Control horizon: Day 0--60 \\

\hline 

$f_d(\cdot), f_c(\cdot)$ \newline (demand and \newline perceived cost \newline functions)
&
$F_k^{\min} =1500 F^{\mathrm{ref}}_k/(\boldsymbol{1}^{\top} F^{\mathrm{ref}})$ in veh/h, $\beta_k=3\,\mathrm{h}^{-1}$, $\gamma=1/3$;
$F^{\mathrm{ref}}$ is the reference OD demand vector from Denver dataset~\cite{xu2024dataset} scaled by $1.3$, $\bbc^{\mathrm{ref}}$ is the reference OD cost vector in~\cite{xu2024dataset}. 
\\ 

\hline

Initial $F(0), \bbc(0)$
& The equilibrium OD-demand and perceived-cost vectors obtained from a 60-day SUMO burn-in under static baseline policy
$\tilde\bbx$ and elastic demand, initialized with $F^{\mathrm{ref}}$ and $\bbc^{\mathrm{ref}}$. \\

\hline

$\ccalZ $ (feasible \newline policy space) 
& Fixed cycle length $90$s, offset $0$s, minimum green time $5$ s, no all-red phase; Junction-specific constant yellow times $\{3, 5, 6\}$ s and number of main phases $\{2,4\}$ set by SUMO.
$x^{\min}_j, x^{\max}_j$ are determined accordingly for all junctions $j\in\{1,\dots,p\}$.
\\

\hline

PSC & $M=50$, $\sigma=0.15$, $\eta_t=0.05$, $w=1.0$, \\
 &  Adam optimizer, 1--10 gradient steps per day. \\
Offline simulator & SUMO, $\rho\in\{1.0,1.2\}$, $\widehat{F}(t) = F(t)\odot \xi$ \\
\bottomrule
\end{tabular}

\end{table}

\begin{figure}[!t]
\vspace{-0.1in}
\centering
\includegraphics[width=0.95\linewidth]{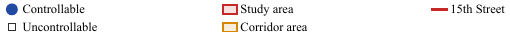}%
\vspace{-0.1in}

\par\noindent
\makebox[\linewidth][c]{%
    \subfloat[Denver, CO]{%
        \includegraphics[width=0.46\linewidth]{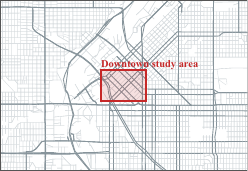}%
        \label{fig:simulation_map_denver}
    }%
    \hspace{0.02\linewidth}%
    \subfloat[Enlarged community map]{%
        \includegraphics[width=0.46\linewidth]{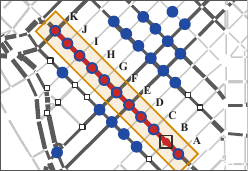}%
        \label{fig:simulation_map_community}
    }%
}
\vspace{-0.1in}
\caption{Simulation maps: (a) Denver with the downtown study area
outlined in red; (b) the enlarged study network with 28 controllable
intersections and the analyzed corridor.
The corridor is northwestbound 15th Street intersects A (Tremont Place) to K (Market Street), with B uncontrollable.
}
\label{fig:simulation_maps}
\end{figure}

\section{Numerical Evaluation}\label{sec:results}
We evaluate day-to-day retiming on a road network in downtown Denver, CO~\cite{xu2024dataset} (see Fig.~\ref{fig:simulation_maps}), using the Simulation of Urban MObility (SUMO)~\cite{lopez2018sumo}. 
The network includes 39 signalized intersections, out of which 28 are controlled, while 11 intersections with irregular phase structures retain their initial SUMO timings to represent partial network controllability. 
The traffic system under control is emulated by SUMO with true demand $F(t)$, online rerouting every 60s, and \texttt{weights.random-factor} set as $\rho=1$.
The detailed configurations are listed in Table~\ref{tab:experiment_settings}.\footnote{Source code will be available at \url{https://github.com/setsunajiang/d2d_retiming}.}

\begin{figure*}[!t]
\centering

\includegraphics[
    width=0.8\textwidth
]{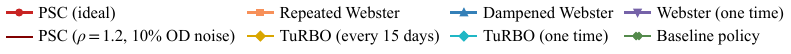}%
\vspace{-0.1in}

\subfloat[TSS]{%
    \includegraphics[
        width=0.24\textwidth
    ]{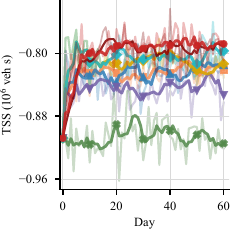}%
    \label{fig:trajectory_noise10_tss}
}
\hfill
\subfloat[Total departed vehicles]{%
    \includegraphics[
        width=0.24\textwidth
    ]{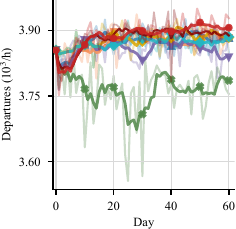}%
    \label{fig:trajectory_noise10_departed}
}
\hfill
\subfloat[P95 cycle-peak queue length]{%
    \includegraphics[
        width=0.24\textwidth
    ]{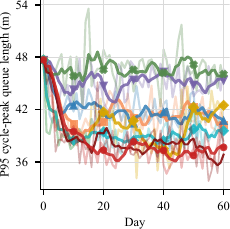}%
    \label{fig:trajectory_noise10_queue}
}
\hfill
\subfloat[Average waiting time per trip]{%
    \includegraphics[
        width=0.24\textwidth
    ]{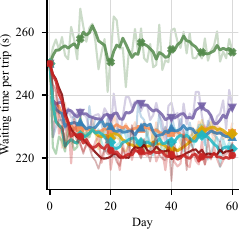}%
    \label{fig:trajectory_noise10_waiting}
}

\caption{Day-to-day retiming control trajectories in downtown Denver, CO. Translucent unmarked lines show daily observations; darker lines show five-day trailing simple moving averages. 
(a) Incremental TSS, (b) Network-wide throughput in departed vehicles per hour,
(c) P95 cycle-peak queue length in meters, averaged within each junction and then across upper-quantile junctions (75-100 percentile), and (d) Network-wide average waiting time per trip.}
\label{fig:trajectory}
\vspace{-0.1in}
\end{figure*}

Seven day-to-day control methods are evaluated with the same initial Day 0 condition $(F(0),\bbc(0),\tilde\bbx)$, yielding different 60-day trajectories of signal-timing policy and OD demand.
These methods are described as follows:

\textit{Baseline policy $\tilde \bbx$:} 
The default static timing plan initialized by SUMO: equal green splits on all  controllable two-phase intersections, with a four-main-phase exception.

\textit{PSC (ideal):}
Our method in Section~\ref{sec:solution} with the parameters in Table~\ref{tab:experiment_settings}. 
The offline simulator is SUMO with noisy OD demand $\widehat F(t)=F(t)\odot\xi$, where $\xi\sim\mathcal N(\bbI,\sigma_F^2\bbI)$, and a mismatched online routing parameter $\rho$, which controls routing uncertainties in SUMO by independently perturbing each edge weight by a random factor drawn from $\mathbb{U}(1,\rho)$. 
We set $(\rho,\sigma_F)$ as $(1,0)$ for \textit{PSC (ideal)}, and as $(1.2,0.05)$ and $(1.2,0.10)$ for two mismatched PSC variants.

\textit{Webster~\cite{webster1958traffic}:} we use Webster approach in three ways: 
\textit{Webster (one time)} obtains a static green-split policy based on day-0 observation and then keeps it for the rest.
\textit{Repeated Webster~\cite{xiao2015combined}} updates the green-split policy $ \bbx^{\mathrm{Webster}}(t) $ based on previous-day observations.
\textit{Dampened Webster} updates the policy as
$\bbx(t)
=
\bigl(1-\kappa_t\bigr)\bbx(t-1)
+
\kappa_t\bbx^{\mathrm{Webster}}(t)$,
where
$\kappa_t=2/(\min\{t,10\}+1)$~\cite{hunter1986}.

\textit{TuRBO:} We reproduce TuRBO according to~\cite{gong2025turbo}, and obtain 1) a static policy for days 1-60 based on $F(0)$, denoted as \textit{TuRBO (one time)}; and 2) a periodically updated policy, denoted as \textit{TuRBO (every 15 days)}, by running TuRBO on days $ t\in\{15, 30, 45\}$ based on the corresponding local OD demand $F(t)$ due to its high computational cost.

\subsection{Performance Metrics}\label{sec:metrics}
\paragraph{Total social surplus}

We instantiate the observed TSS $\widehat J$ in \eqref{E:observed_TSS} with completed SUMO trips. The total travel time of trip $i$ is $T_i=T_i^{\mathrm{ff}} + L_i$, where $T_i^{\mathrm{ff}}$ and $L_i$ are its free-flow travel time and \texttt{timeLoss}. For OD pair $k$, let
\[
\bar T_k^{\mathrm{ff}}
=
\frac{1}{|\ccalV_k^{\mathrm{c}}|}
\sum_{i\in\ccalV_k^{\mathrm{c}}}T_i^{\mathrm{ff}},
\qquad
\bar L_k
=
\frac{1}{|\ccalV_k^{\mathrm{c}}|}
\sum_{i\in\ccalV_k^{\mathrm{c}}}L_i .
\]
We set $b_k=(1+\alpha)\bar T_k^{\mathrm{ff}}$ and $C_k=\bar T_k^{\mathrm{ff}}+\bar L_k$, yielding the average surplus $b_k-C_k=\alpha\bar T_k^{\mathrm{ff}}-\bar L_k$. 
We set the relative benefit factor of an automobile trip as $\alpha=1$.

\paragraph{P95 queue length}

We collect the 95th percentile cycle-peak queue length of an approach $a$ in day $t$ as:
\[
Q^{P95}_{a}(t) = \operatorname{P}_{95}\left(\{Q_{a,r}^{\max}\}_{r\in t}\right),\quad Q_{a,r}^{\max} = \max_{\tau\in r} Q_a(\tau),
\]
where $Q_a(\tau)$ is the maximum E2 jam length in meters among the lanes of approach $a$, and $ Q_{a,r}^{\max} $ is its peak value in cycle $r$. 
The per-junction $Q^{P95}_{j}(t)$ is the average $Q^{P95}_{a}(t)$ of its incoming approaches, and we report the daily average of junctions with $Q^{P95}_{j}(t)$ at or above the 75th percentile.

\begin{figure}[!t]
\centering
\subfloat[Policy trajectories]{%
    \includegraphics[
        width=0.48\linewidth
    ]{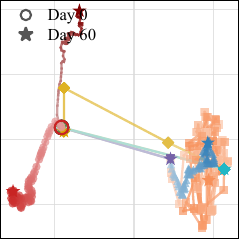}%
    \label{fig:policy_tsne}
}%
\hfill
\subfloat[OD-demand trajectories]{%
    \includegraphics[
        width=0.48\linewidth
    ]{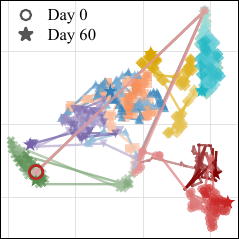}%
    \label{fig:od_tsne}
}
\\

\subfloat[Five-day policy distance]{%
    \includegraphics[
        width=0.48\linewidth
    ]{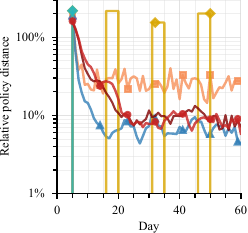}%
    \label{fig:policy_distance_5day}
}%
\hfill
\subfloat[Five-day OD-demand distance]{%
    \includegraphics[
        width=0.48\linewidth
    ]{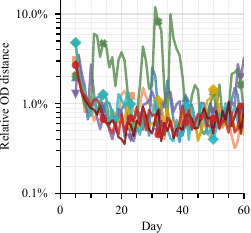}%
    \label{fig:od_distance_5day}
}

\caption{The evolutions of policy and demand under the seven controllers.
(a)--(b) 2D t-SNE projections of the signal-timing
policies and OD-demand vectors, where darker colors indicate later
days. (c)--(d) Relative Euclidean distances between Days $t$ and
$t-5$ for the normalized green-split policies and OD-demand vectors. 
Lower distances indicate smaller
five-day changes. Method colors and marker shapes follow
Fig.~\ref{fig:trajectory}.}
\label{fig:policy_demand_stability}
\end{figure}

\paragraph{Average waiting time}
The network-wide average waiting time per trip is the total time of all vehicles traveling slower than $0.1\,\mathrm{m/s}$ divided by the total departures, during the $3600$s main simulation period.
Intersection-level $\widehat\bbw_j$ is obtained by total waiting time on all of its incoming approaches, divided by their total number of \texttt{entered} and \texttt{departed} vehicle events in the main period.

\subsection{Day-to-Day Control Trajectories}
We report the realized trajectories of network-wide TSS, throughput, upper-quantile P95 queue length, and average waiting time per trip under the seven controllers in Figs.~\ref{fig:trajectory_noise10_tss}-\ref{fig:trajectory_noise10_waiting}.
Their realized trajectories of high-dimensional policies and OD demands are illustrated in Figs.~\ref{fig:policy_demand_stability} via 2D t-SNE projections and 5-day relative Euclidean distances.
All methods improve traffic performances over the baseline policy.

The first observation is that static timing policy can be unstable, as shown by the large fluctuations in the trajectories of the baseline policy and Webster (one time), especially in TSS, throughput, and 5-day relative OD distance.
TuRBO (one time) is more stable than them but its OD demand trajectory still drift substantially (Fig.~\ref{fig:od_tsne}) with higher fluctuations than PSC in 5-day OD distance (Fig.~\ref{fig:od_distance_5day}).

Repeated and Dampened Webster methods achieve similar TSS, throughput, queue length, and average waiting time, despite that their daily policy fluctuations differ significantly (Figs.~\ref{fig:policy_tsne} and \ref{fig:policy_distance_5day}). 
Compared to Repeated Webster, the seemingly more stable policy of Dampened Webster results in larger fluctuations in OD demand (\ref{fig:od_distance_5day}).
TuRBO (every 15 days) performs worse than TuRBO (one time) in both performance and stability, demonstrating the limits of repeatedly running end-state optimum methods.

The two PSC variants (ideal, $\rho=1,\sigma_F=0$) and ($\rho=1.2, \sigma_F=0.1$) perform similarly, achieving the best performance across all metrics in Figs.~\ref{fig:trajectory}.
Compared to others, PSC can guide the policy and demand to co-evolve to more stable regions based on~\eqref{E:pstable-neighborhood}, evidenced by simultaneously lower 5-day distances of policy (Fig.~\ref{fig:policy_distance_5day}) and OD demand (Fig.~\ref{fig:od_distance_5day}), as well as smaller regions of drifting in their projected trajectories in Figs.~\ref{fig:policy_tsne} and~\ref{fig:od_tsne}.

\subsection{Intersection-wise Level of Service}
\begin{figure}[!t]
\centering
\includegraphics[width=0.85\linewidth]{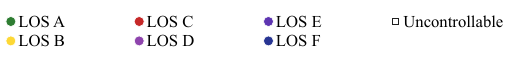}%
\vspace{-0.05in}

\includegraphics[width=0.9\linewidth]{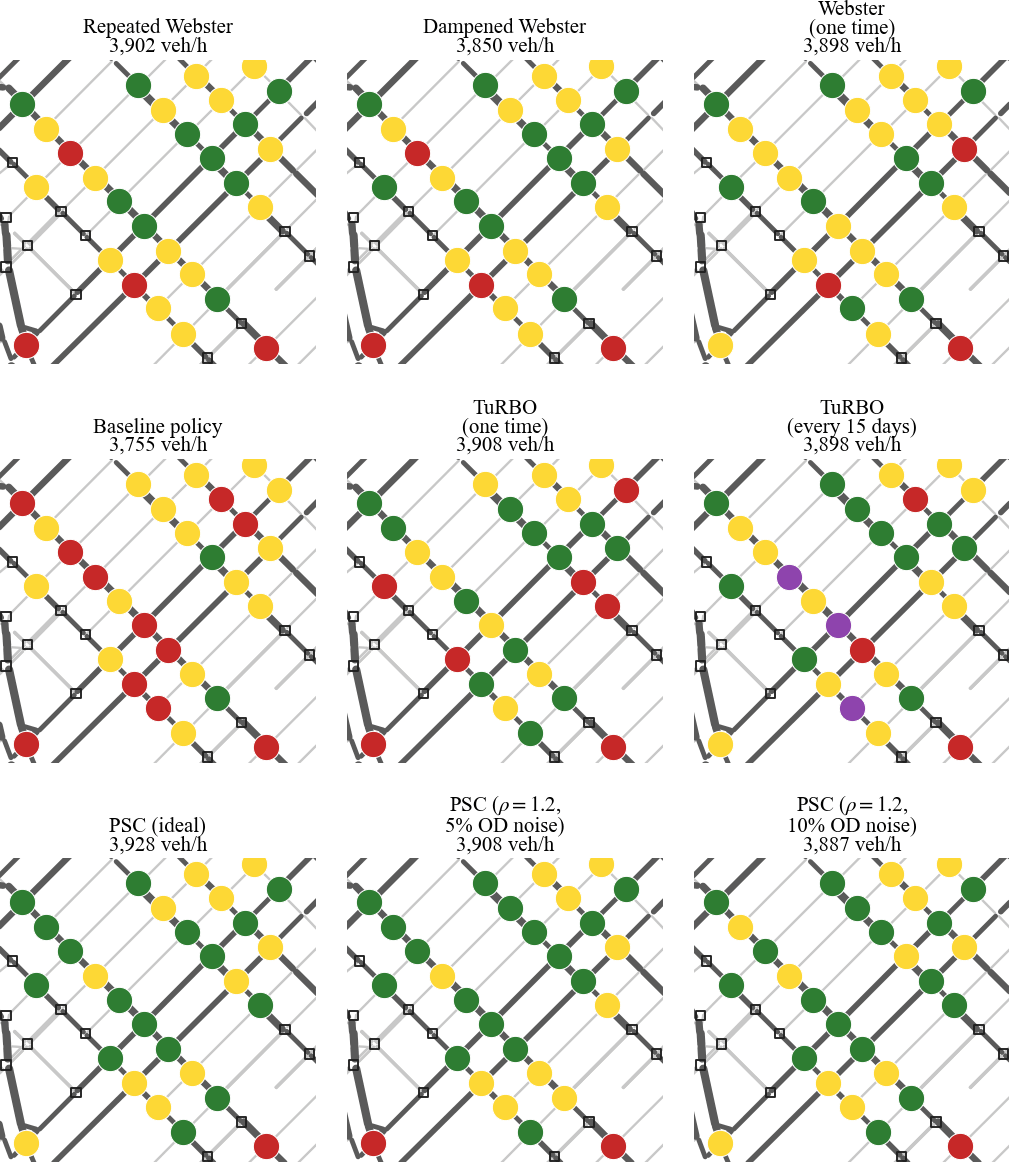}

\caption{Junction-level LOS comparison across nine control methods
on Day 60, using each method's own resulting demand and control policy. Each panel reports network throughput in vehicles per hour.}
\label{fig:los_day60_own_demand}
\end{figure}

Fig.~\ref{fig:los_day60_own_demand} presents the snapshots of junction-wise Level of Service (LOS) and realized total throughput under different controllers on the last day of their control trajectories. 
They illustrate the traffic performance of signal-timing policies spatially. 
For each controllable junction, we define a control-delay proxy as the total SUMO \texttt{timeLoss} accumulated on its incoming lanes divided by the total number of \texttt{entered} and \texttt{departed} vehicle events on those lanes. LOS A--F correspond respectively to delay ranges $[0,10]$, $(10,20]$, $(20,35]$, $(35,55]$, $(55,80]$, and $(80,\infty)$ seconds per vehicle.

The PSC (ideal) achieves the highest throughput, with 16 of the 28 controllable junctions operating at LOS A, 11 at LOS B, and only one at LOS C, outperforming others. 
The two mismatched PSC variants also outperform non-PSC methods in LOS, with 16 LOS A junctions, while their throughput is slightly lower.

TuRBO (one time) achieves 2-nd highest throughput with 11 LOS A and 7 LOS C junctions, whereas TuRBO (every 15 days) causes deterioration in both throughput and LOS, with 3 junctions deteriorate to LOS D. 
Compared to TuRBO (one time), the three Webster controllers achieve slightly lower throughput but less congested junctions (LOS C or worse).

\begin{table*}[!t]
\centering
\caption{Network-wide performance realized by day-50 policies under three common demand profiles. Best values are bold.}
\label{tab:demand_profile_performance}
\vspace{-0.04in}

\setlength{\tabcolsep}{3pt}
\renewcommand{\arraystretch}{1.08}
\footnotesize

\begin{tabular*}{\textwidth}{
@{\extracolsep{\fill}}
l
rrr
rrr
rrr
}
\toprule
& \multicolumn{3}{c}{Reference profile: $F(0)$}
& \multicolumn{3}{c}{Light-load profile:
$0.5F^{\mathrm{PSC}}(50)$}
& \multicolumn{3}{c}{Heavy-load profile:
$2F^{\mathrm{PSC}}(50)$} \\
\cmidrule(lr){2-4}
\cmidrule(lr){5-7}
\cmidrule(lr){8-10}
Method
& TSS ($10^3$) & Thpt. (veh/h) & Wait. (s)
& TSS ($10^3$) & Thpt. (veh/h) & Wait. (s)
& TSS ($10^3$) & Thpt. (veh/h) & Wait. (s) \\
\midrule

Baseline policy
& $-874.8$ & 3769 & 248.7
& $-550.3$ & 2535 & 237.1
& $-1943.7$ & 5711 & 339.8 \\

Webster (one time)
& $-803.0$ & 3819 & 228.5
& $-485.9$ & 2563 & 214.5
& $-1675.2$ & 5845 & 291.1 \\

Repeated Webster
& $-809.8$ & 3812 & 230.1
& $-449.2$ & 2567 & 205.7
& $-1633.5$ & 5990 & 271.6 \\

Dampened Webster
& $-768.6$ & \textbf{3838} & 219.3
& $-448.2$ & 2569 & 203.4
& $-1705.1$ & 5970 & 282.6 \\

TuRBO (one time)
& $-777.2$ & 3790 & 225.4
& $-447.6$ & 2526 & 207.5
& $-1625.9$ & \textbf{5992} & 272.1 \\

TuRBO (every 15 days)
& $-790.7$ & 3816 & 223.7
& $-414.4$ & \textbf{2572} & 195.6
& $-1767.0$ & 5934 & 284.6 \\
\hline 
PSC (ideal)
& $-754.6$ & 3823 & 218.8
& $\mathbf{-383.6}$ & \textbf{2572} & \textbf{184.4}
& $-1553.3$ & 5977 & 263.5 \\

PSC ($\rho=1.2$, $\sigma_F=5\%$)
& $\mathbf{-749.9}$ & 3778 & \textbf{217.5}
& $-418.6$ & 2557 & 196.3
& $\mathbf{-1547.8}$ & 5991 & 263.8 \\

PSC ($\rho=1.2$, $\sigma_F=10\%$)
& $-770.9$ & 3833 & 221.4
& $-414.2$ & 2532 & 196.5
& $-1560.1$ & 5969 & \textbf{262.2} \\

\bottomrule
\end{tabular*}
\end{table*}

\begin{figure*}[!t]
\centering
\includegraphics[width=0.8\textwidth]{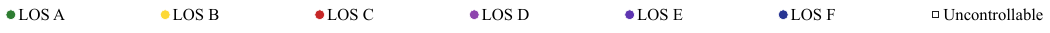}%
\vspace{-0.03in}

\subfloat[Reference profile: $F(0)$]{%
    \includegraphics[width=0.3\textwidth]
    {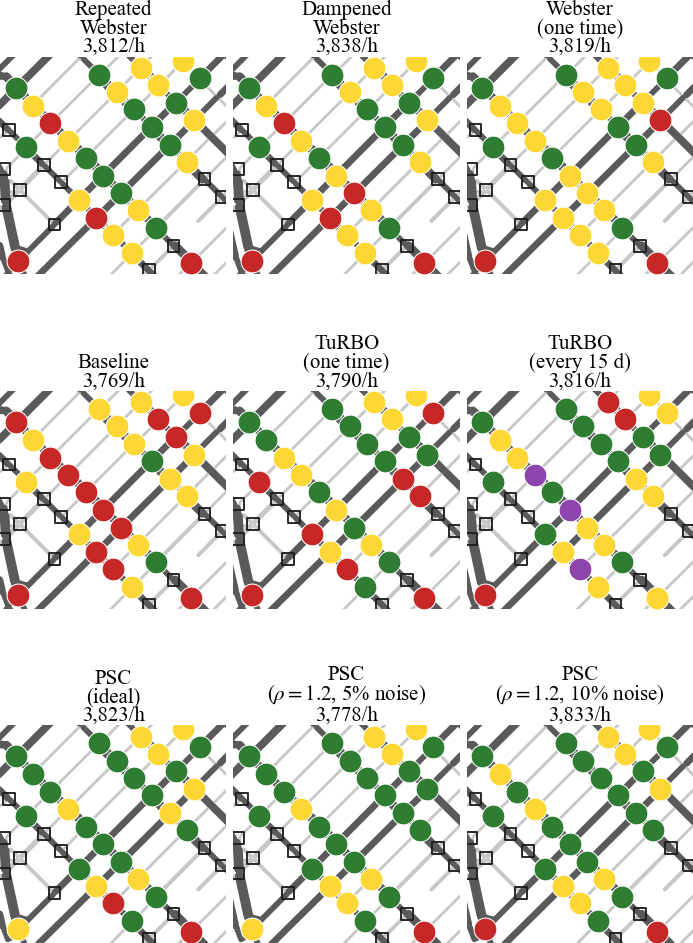}%
    \label{fig:los_day50_common_day0}
}\hfill
\subfloat[Light-load profile: $0.5F^{\mathrm{PSC}}(50)$]{%
    \includegraphics[width=0.3\textwidth]
    {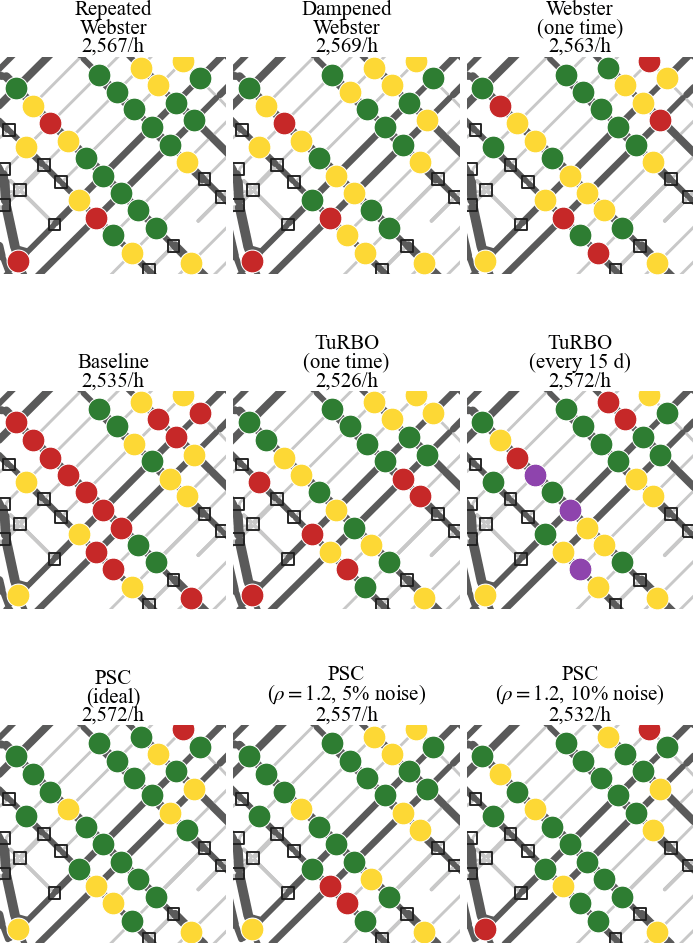}%
    \label{fig:los_day50_demand_x05}
}\hfill
\subfloat[Heavy-load profile: $2F^{\mathrm{PSC}}(50)$]{%
    \includegraphics[width=0.3\textwidth]
    {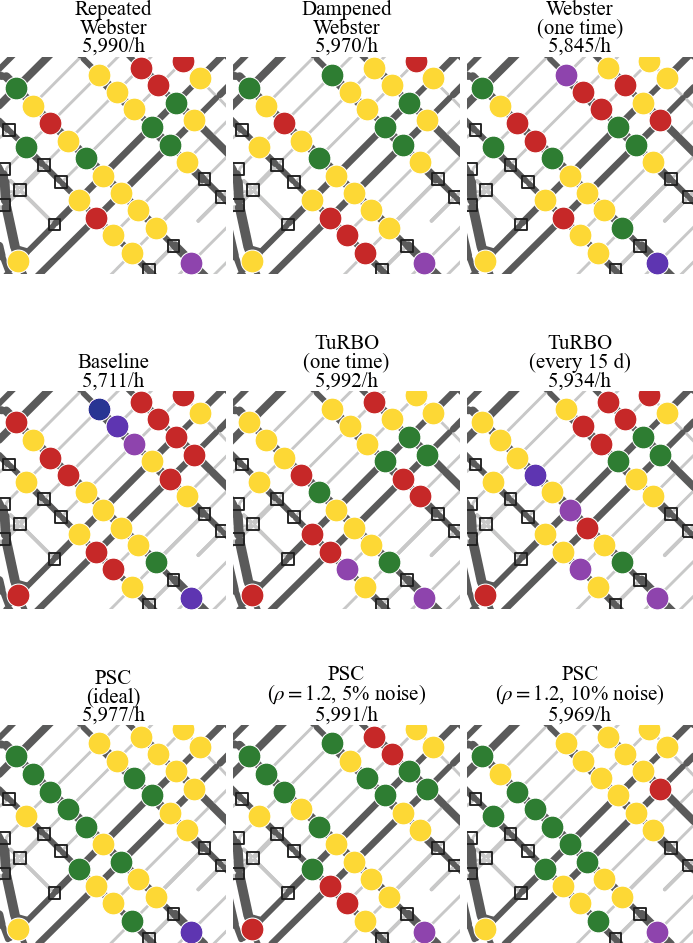}%
    \label{fig:los_day50_demand_x2}
}

\caption{Realized per-junction LOS and network throughput under nine
fixed Day 50 control policies and three common demand profiles:
(a) the reference profile, given by the Day 0 equilibrium demand
$F(0)$; (b) the light-load profile,
$0.5F^{\mathrm{PSC}}(50)$; and (c) the heavy-load profile,
$2F^{\mathrm{PSC}}(50)$. The light- and heavy-load profiles are
obtained by uniformly scaling PSC (ideal) Day-50 OD vector.
Within each subfigure, all nine policies are evaluated using the same
OD demand. The demands are not method-specific.
}
\label{fig:los_day50_stress}
\end{figure*}

\begin{figure*}[!t]
    \centering
    \includegraphics[
        width=\textwidth
    ]{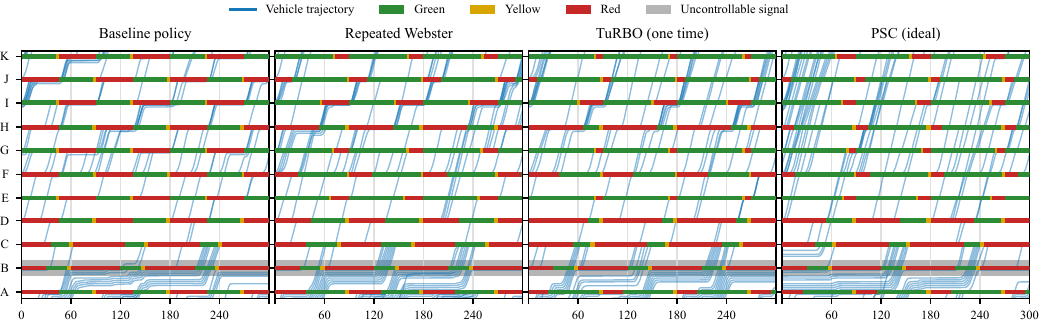}
    \caption{Time--space diagrams for northwestbound traffic on 15th
    Street in Denver, highlighted in Fig.~\ref{fig:simulation_map_community}, under the common Day 50 OD demand from the PSC (ideal) trajectory. From left to right, the panels show the static baseline, the Day 50 Repeated Webster policy, TuRBO (one time), and the Day 50 PSC (ideal) policy. All panels use the same SUMO interval from 2700 to 3000 s; the horizontal axis reports elapsed time since 2700 s.}
    \label{fig:time_space_15th_street}
\end{figure*}

\subsection{Generalization across Demand Profiles}

To test how well the resulting signal-timing policies generalize, we test the day-50 policies $\bbx(50)$ from different controllers under three OD demand profiles:
1) The \textit{reference profile} uses the common Day-0 OD demand $F(0)$. 
2) The \emph{light-load profile}, $0.5F^{\mathrm{PSC}}(50)$, 
and 3) the \emph{heavy-load profile}, $2F^{\mathrm{PSC}}(50)$, where $ F^{\mathrm{PSC}}(50) $ is the day-50 OD demand from the demand trajectory of PSC (ideal).

Table~\ref{tab:demand_profile_performance} and
Fig.~\ref{fig:los_day50_stress} present the network-wide performances and intersection-wise LOS results. 
Under these three demand profiles, the three PSC day-50 policies achieve the best TSS and LOS, and perform consistently well in throughput and waiting time. 
In particular, the LOS levels under the three PSC policies exhibit least amount of congested intersections across different traffic load levels. 
Compared to Webster day-50 policies, the two TuRBO day-50 policies generally cause more localized congestion across three demand profiles.

\subsection{Corridor-Level Microscopic Analysis}

Figure~\ref{fig:time_space_15th_street} provides a microscopic view of vehicle progression along the 15th Street corridor highlighted in Fig.~\ref{fig:simulation_map_community}. 
The baseline policy causes the most frequent stops as vehicles travel downstream. 
Repeated Webster and TuRBO (one time) only brings partial improvement in some segments of the corridor. 
PSC aligns more trajectories with successive green intervals, particularly from C to K. 
Such smoother progression helps explain the lower congestion under PSC.

\section{Conclusion and Future Work}

In this work, we propose a gradient-based incremental approach for day-to-day fixed-time signal retiming when travelers adapt both their route choices and travel demand to the deployed policy.
Numerical experiments show that the proposed method produces more stable policy--demand trajectories, improves network-wide and junction-level traffic performances and remains effective under demand-estimation error, simulator routing mismatch, and across substantially different demand profiles.
These results demonstrate that gradient-based gradual policy update can guide the co-evolving signal timing policy and travelers' adaptation toward performative stability while achieving favorable traffic performance.

The current zeroth-order optimization framework does not explicitly exploit spatial dependencies among intersections or model the long-run induced-demand effect required for performative optimality.
Future work will investigate gray-box and spatial surrogate models to improve policy sample efficiency and scalability, behavioral and equilibrium models for performative optimal control, and broader signal-timing policies including cycle length, offsets, and more general intersection configurations.

\bibliographystyle{IEEEtran}
\bibliography{sj_ref,traffics}

\end{document}